\documentclass[final]{aipproc}
\usepackage{graphicx}
\usepackage{amsmath}
\usepackage{amssymb}
\layoutstyle{6x9}

\begin{document}

\title{Charmonium in Medium and at RHIC}


\author{D. Blaschke}{
  address={GSI Darmstadt, Planckstr. 1, D-64291 Darmstadt, Germany}
,altaddress={BLTP @ JINR Dubna, Joliot-Curie Str.\ 6, 141980 Dubna, Russia} 
}

\author{V.L. Yudichev}{
  address={BLTP @ JINR Dubna, Joliot-Curie Str.\ 6, 141980 Dubna, Russia} 
}

\begin{abstract}
The production of charmonia in relativistic heavy-ion collisions is
formulated within the framework of a quantum kinetic theory simultaneously
accounting for breakup and recombination processes.
The survival probability of $J/\psi$ mesons is evaluated for
a Mott-Hagedorn resonance gas, whereby
the pattern of anomalous $J/\psi$ suppression at CERN-SPS
can be explained by a strong coupling to open-charm states and their
in-medium modification.
Recent data for charmonium suppression
from $Au$-$Au$ collisions at RHIC are discussed.
\end{abstract}

\maketitle


\section{Introduction}
Recent experiments at RHIC have revealed that the quark-gluon
plasma in the vicinity of the QCD phase transition is strongly
correlated (sQGP) with a low viscosity (perfect liquid), 
see \cite{Muller:2006} for a recent review. It has
been conjectured that this result of a flow data analysis might be
due to the presence of hadronic bound states \cite{Shuryak:2004kh}
or resonances \cite{Blaschke:2005cm} above the critical
temperature $T_c$ for the phase transition. The persistence of
hadronic correlations above $T_c$ has been obtained before in
lattice QCD simulations. Since the steep increase of the effective
number of degrees of freedom which defines $T_c$ as seen in
lattice data for the energy density $\varepsilon/T^4$ could nicely
be understood in terms of the Hagedorn resonance gas
\cite{Karsch:2003vd}, the question appeared whether the QCD
thermodynamics even above  $T_c$ could be interpreted in terms of
hadronic degrees of freedom.
Assuming a spectral broadening of the hadrons due to a Mott transition an
alternative interpretation of the lattice data has been given
\cite{Blaschke:2004xg}.
The key element of a microscopic formulation of such a Mott-Hagedorn resonance
gas is the description of the Mott
transition for hadrons in hot and dense matter in terms of the hadronic
spectral functions.
First steps in this direction and possible consequences for the behavior of
charmonium states in the sQGP at RHIC will be discussed in this contribution.
\section{Charmonium kinetics in the sQGP}
According to the above picture of the sQGP as being composed of strong
hadronic correlations at a large phase space density, a huge number of
energetically degenerate decompositions in terms of a hadronic basis emerges
which are related to each other by flipping the string-type interaction links
between quark-antiquark pairs.
The left two panels of Fig.~1 give an example for such a string-flip process
actually involving four pairs.
At each time step the system has
strong nearest neighbor correlations resulting in a liquid-like
 pair correlation function, see Fig. 1 (right panel).
\begin{figure}
  \includegraphics[height=.22\textheight]{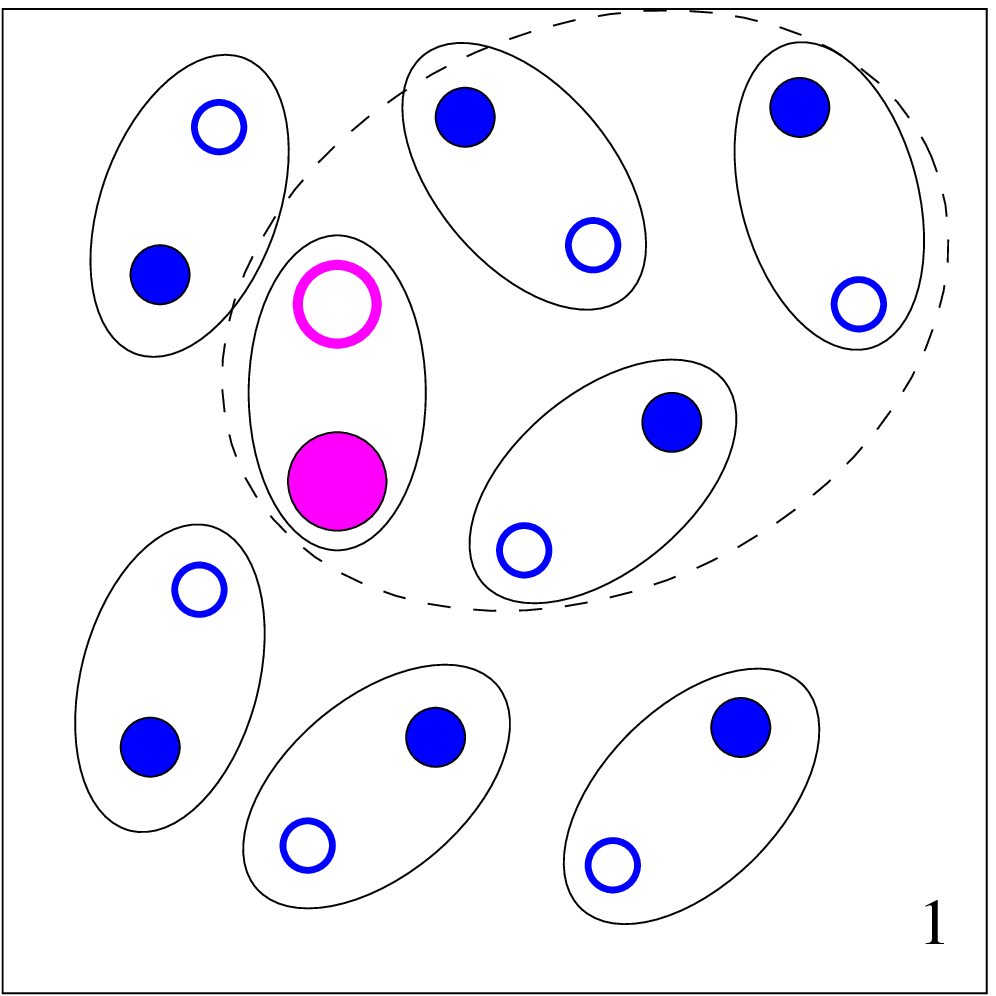}
\includegraphics[height=.22\textheight]{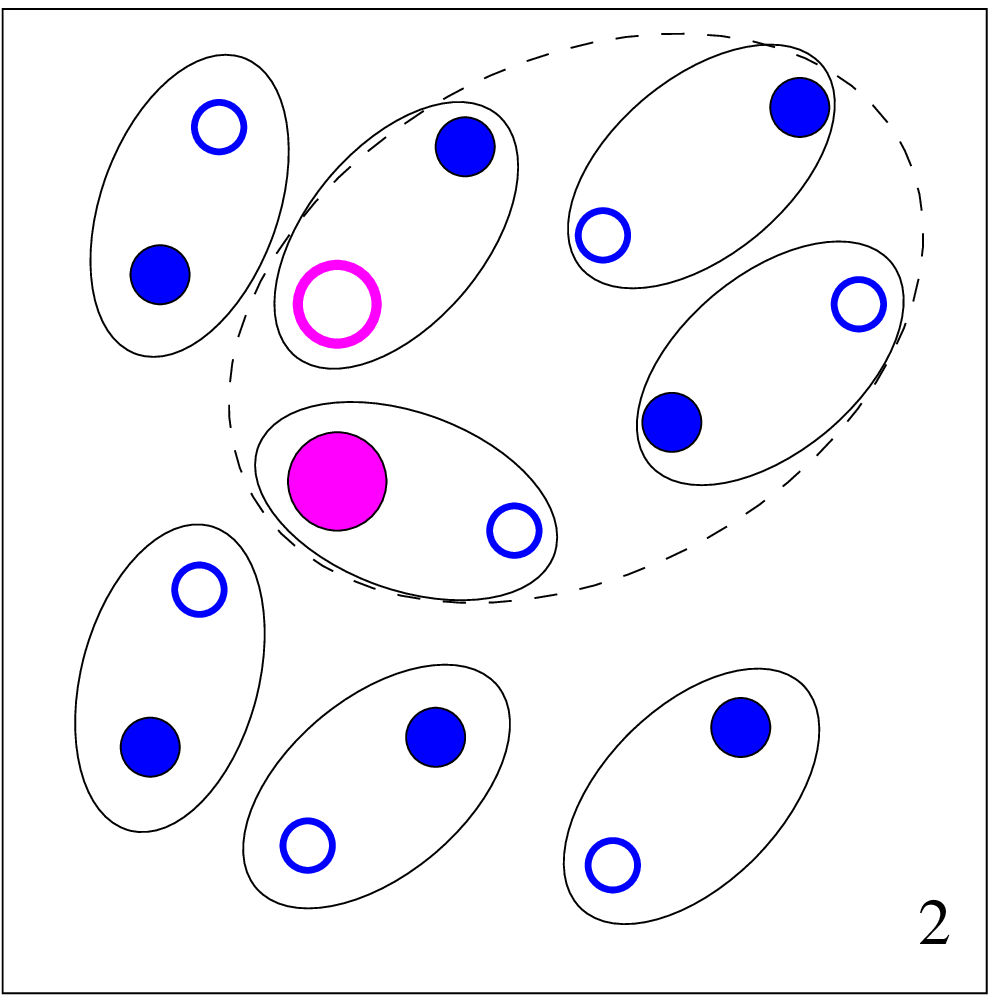}
\includegraphics[height=.22\textheight]{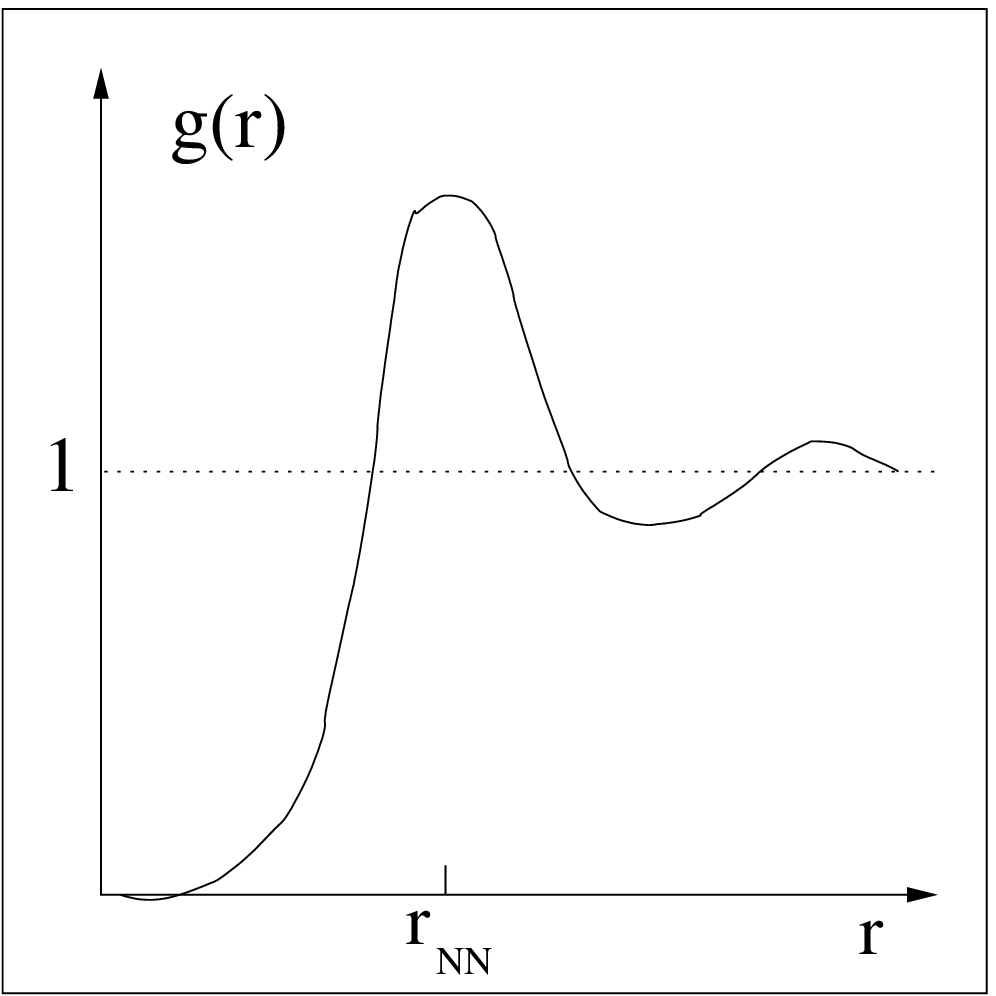}
  \caption{Hadronic correlations  in quark matter depicted by encircling pairs
of quarks (full dots) and antiquarks (open dots).
A rearrangement collision (string-flip process) relates the left and middle
panels.
The right panel illustrates the liquid-like pair correlation function with a
nearest-neighbor peak.}
\end{figure}
The string-flip processes may occur at a high rate leading to a sizeable
spectral broadening of all hadronic basis states in accordance with the
Mott-\-Hagedorn resonance gas picture which provides a physical understanding
for phenomena like fast chemical equilibration and anomalous J/$\psi$
suppression.
The suppression of the production of charmonia states at increasing the
energy density in a heavy ion collision has been suggested as an indicator for
QGP formation 
on the basis of a simple argument: the charmonia states obtained
as solutions of the two-particle Schr\"odinger equation for a
temperature-\- dependent heavy-quark potential from lattice-QCD
show a Mott-transition of the J/$\psi$ bound state to the
continuum of scattering states at $T\sim 1.2~ T_c$, see
\cite{Blaschke:2005jg} and Refs.\ therein. This picture, however,
is not appropriate for the situation of the sQGP where the medium
consists of strongly correlated quarks in hadronic resonance
states. The dominant process leading to J/$\psi$ suppression is
then a quark rearrangement collision (string-flip) process
\cite{Blaschke:1992qa,Martins:1994hd} corresponding to charmonium
dissociation into open charm states, see Fig.~2. It has been shown
that this process, which in the vacuum and at temperatures below
$T_c$ is suppressed by a reaction threshold of $\sim 600$ MeV,
becomes critically enhanced at the chiral phase transition since
the open charm mesons undergo a Mott-like transition with a
spectral broadening effectively lowering the charm dissociation
threshold \cite{Burau:2000pn,Blaschke:2003ji}.
The proper quantum kinetic formulation of the problem of heavy quarkonia 
states
\begin{figure}[h!]
  \includegraphics[height=.2\textheight]{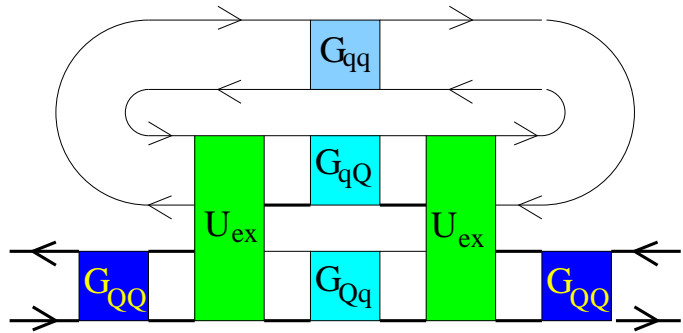}
  \caption{Quark rearrangement (string-flip) couples heavy and light
quarkonia to heavy mesons.}
\end{figure}
in the sQGP can be given along the lines of the Kadanoff-Baym equation
approach, where the propagators of the composite states in the quark
rearrangement process are obtained from solutions of the Bethe-Salpeter
equations for the corresponding T-matrices \cite{Mannarelli:2005pz}, see
Fig.~2.
The momentum-dependent  lifetime $\tau_{Q\bar Q}(p)$ of the charmonia states
is given by
\begin{eqnarray}
\tau^{-1}_{Q\bar Q}(p) &=&  \Sigma^>_{Q\bar Q}(p) - \Sigma^<_{Q\bar Q}(p)~,\\
\Sigma^{\stackrel{>}{<}}_{Q\bar Q}(p) &=& \int\limits_{p'}
\int\limits_{p_1}\int\limits_{p_2} (2\pi)^4 \delta_{p,p';p_1,p_2}
\left|{U}_{ex}\right|^2
G^{\stackrel{<}{>}}_{q\bar q}(p')~
G^{\stackrel{>}{<}}_{q\bar Q}(p_1)~
G^{\stackrel{>}{<}}_{Q\bar q}(p_2)~,
\end{eqnarray}
where $\Sigma^>_{Q\bar Q}(p)$ ($\Sigma^<_{Q\bar Q}(p)$) stand for the effect
of loss (gain) processes and depend on the hadronic Green functions
$G^{>}_{i}(p) = [1 + f_i(p)] A_i(p)$ and $G^{<}_{i}(p) = f_i(p) A_i(p)$
defined by their spectral functions
$A_i$ and phase space occupation functions $f_i$, 
where $i={q\bar Q}, {Q\bar q}, {q\bar q}$.
Within this approach the observed saturation of the anomalous J/$\psi$
suppression in recent RHIC experiments is explained by the
detailed balance between gain and loss processes when the charm meson
phase space is sufficiently populated.
\section{Conclusions}
A microscopic formulation of the QCD phase transition is suggested in terms
of a hadronic basis with the Hagedorn mass spectrum and a spectral broadening
due to frequent quark rearrangement processes in a hot and dense medium
(Mott dissociation).
the approach provides a simultaneous understanding of Lattice QCD
thermodynamics, hadronic correlations above $T_c$, anomalous J/$\psi$
suppression at CERN-SPS and its saturation at RHIC experiments.
\begin{theacknowledgments}
We acknowledge discussions with H. van Hees, O. Kaczmarek, Yu. Kalinovsky, 
E. Laermann and R. Rapp. V.L.Yu. was supported by DFG under grant No.\ 436 RUS
17/128/03 and by the Helmholtz International Summer School program
(HISS Dubna).
\end{theacknowledgments}

\end{document}